\documentclass[twocolumn]{aastex6} 
\usepackage[utf8]{inputenc}
\usepackage{graphicx}
\usepackage{epstopdf}
\usepackage{natbib}
\usepackage{rotating}
\usepackage{makecell}
\usepackage{amsmath}
\usepackage{morefloats} 
\usepackage{hyperref}

\def\bea{\begin{eqnarray}}
\def\eea{\end{eqnarray}}
 
\newcommand{\HI}{\hbox{{\rm H}\kern 0.1em{\sc i}}}
\newcommand{\HII}{\hbox{{\rm H}\kern 0.1em{\sc ii}}}
\newcommand{\NII}{\hbox{{\rm N}\kern 0.1em{\sc ii}}}

\accepted{in ApJ May 2023}

\begin{document}

\title{Circumgalactic Ly$\alpha$ Nebulae in Overdense Quasar Pair Regions Observed with the Palomar Cosmic Web Imager}

\author{Jessica S. Li\altaffilmark{1}, Carlos J. Vargas\altaffilmark{1}, Donal O'Sullivan\altaffilmark{2}, Erika Hamden\altaffilmark{1}, Zheng Cai\altaffilmark{3,4}, Mateusz Matuszewski\altaffilmark{2}, Christopher Martin\altaffilmark{2}, Miriam Keppler\altaffilmark{1}, Haeun Chung \altaffilmark{1}, Nicole Melso \altaffilmark{1}, Shiwu Zhang \altaffilmark{3}}

\altaffiltext{1}{Department of Astronomy and Steward Observatory, University of Arizona, Tucson, AZ, USA}\altaffiltext{2}{Cahill Center for Astrophysics, California Institute of Technology, 1216 East California Boulevard, Mail code 278-17, Pasadena, CA 91125, USA}
\altaffiltext{3}{Department of Astronomy, Tsinghua University, Beijing 100084, China; zcai@mail.tsinghua.edu.cn}
\altaffiltext{4}{UCO/Lick Observatory, University of California, 1156 High Street, Santa Cruz, CA 95064, USA}

\begin{abstract}

The recent discovery of enormous Ly$\alpha$ nebulae (ELANe), characterized by physical extents $>200$ kpc and Ly$\alpha$ luminosities $>10^{44}$ erg s$^{-1}$, provide a unique opportunity to study the intergalactic and circumgalactic medium (IGM/CGM) in distant galaxies. Many existing ELANe detections are associated with local overdensities of active galactic nuclei (AGN). We have initiated a search for ELANe around regions containing pairs of quasi-stellar objects (QSOs) using the Palomar Cosmic Web Imager (PCWI). The first study of this search, \citet{caietal18}, presented results of ELAN0101+0201 which was associated with a QSO pair at $z=2.45$.  In this study, all targets residing in QSO pair environments analyzed have Ly$\alpha$ detections, but only one of the four targets meets the classification criteria of an ELANe associated with a QSO pair region (z$\sim2.87$). The other three sample detections of Ly$\alpha$ nebulae do not meet the size and luminosity criteria to be classified as ELANe.
We find kinematic evidence that the ELANe J1613, is possibly powered {mostly by AGN outflows.} The analysis of circularly-averaged surface brightness profiles of emission from the Ly$\alpha$ regions show that the {Ly$\alpha$ emission around $z\sim2$ QSO pairs is consistent with emission around individual QSOs at $z\sim2$ \citep{caietal19}, which is fainter than that around $z\sim3$ QSOs \citep{battaia2019b,caietal19,fossati2021}}. A larger sample of Ly$\alpha$ at z$\sim$2 will be needed to determine if there is evidence of redshift evolution when compared to nebular emissions at z$\sim$3 from other studies.

\end{abstract}
\section{Introduction}\label{sec:intro}

The intergalactic medium (IGM) and circumgalactic medium (CGM) are of crucial importance in regulating galaxy formation and evolution. They act as reservoirs for fueling star formation in galaxies, and are subject to feedback processes responsible for the ejection of matter from galaxies. Their morphology and kinematics provide stringent constraints to models of galaxy formation \citep{2011Stewart,2015Martin}. In particular, giant Ly$\alpha$ nebulae (also known as ``Ly$\alpha$" blobs, LABs) are characterized by a high luminosity of  Ly$\alpha$ line emission (L$_{\rm{Ly\alpha}}\gtrsim$ $10^{43}$ erg s$^{-1}$), and a spatially large  Ly$\alpha$ emitting region up to hundreds of kpc \citep{2005Dey}.  For example, \citet{matsuda2004} reports the detection of 35 LABs at z = 3.1, two of which are extremely Ly$\alpha$ bright and have large physical extents \citep{steidel2000}.  In another study, \citet{yang2009} discovers four new LABs, where two of the four are powered by QSOs at z = 2.3.  A broadband survey for Ly$\alpha$ nebulae by \citet{prescott2013} confirmed the presence of four large Ly$\alpha$ nebulae, with Ly$\alpha$ luminosities ranging from $\sim$ 10$^{43}$ to 10$^{44}$ erg s$^{-1}$ at z $\sim$ 2-3. A subset of LABs, referred to as enormous Ly$\alpha$ nebulae (ELANe; \citealt{caietal17a,arrigonietal18}), are defined as having extents $>200$ kpc and L$_{\rm{Ly\alpha}}>10^{44}$ erg s$^{-1}$.

The Ly$\alpha$ line is the primary coolant of cold $T \sim10^4$ K low metallicity gas and can be used to map the CGM/IGM via emission. Several mechanisms have been identified that should lead to Ly$\alpha$ emission from the CGM: cooling, in-falling gravitationally heated gas \citep{2000Haiman}, cooling following superwind-driven shocks \citep{2000Taniguchi}, and Ly$\alpha$ fluorescence induced by exposure to UV radiation. Each of these mechanisms is likely to be dominant at different scales, with multiple processes responsible for powering a single Ly$\alpha$ nebula. 

Ly$\alpha$ fluorescence caused by the much stronger UV radiation from a nearby source can also boost emission from a Ly$\alpha$ nebula into the detectable regime \citep{1988Rees,2001Haiman,2010Kollmeier}. The surface brightness of an optically thin Ly$\alpha$ nebulae is roughly proportional to $M_{\rm{gas}}^2$. Strongly clustered, ultra-luminous QSO groups provide us with both conditions: (1) a strong meta-galactic ionising flux from an over-density of AGN; (2) a significant gas overdensity in this system. These strongly clustered groups are the best sites to initially search for giant Ly$\alpha$ nebulae, where Ly$\alpha$ emission can reach IGM-scale distances of $\gtrsim500$ kpc. 

\subsection{Confirmed LABs and ELANe}

Several systematic efforts have been made to search for LABs and ELANe at $z\approx2$. Using the VLT/X-shooter spectrograph, \citet{zafar2011} measures a LAB with a velocity gradient that may indicate gas inflow or outflow that is associated with a QSO pair. Through deep narrow-band imaging surveys around ultra-luminous type-I QSOs, \citet{2014Cantalupo} and \citet{2015Hennawi} discovered two different giant Ly$\alpha$ nebulae with spatial extent $\gtrsim400$ kpc. The Slug nebula \citep{2014Cantalupo} is the first LAB that extends to IGM scales of 500 kpc. It is associated with a QSO pair consisting of one bright and one faint QSO. The Jackpot nebula \citep{2015Hennawi} resides in a massive overdensity of 4 AGN. These two nebulae have similar sizes and surfaces brightnesses (SBs) although QSO luminosities differ by an order of magnitude.  This suggests that the QSOs sufficiently photoionized the surrounding gas resulting in optically thin ELANe. Besides AGN photo-ionization, additional mechanisms, e.g., cooling flows, may play a role in powering Ly$\alpha$ emission in massive halos  \citep{arrigonietal19}. Using the Palomar Cosmic Web Imager (PCWI), \citet{2015Martin} reveals large velocity shifts of a giant LAB, the Slug Nebula.  Assuming these shifts are due to kinematics, these PCWI observations yielded the first 3-D picture of how massive galaxies acquire gas from the IGM. The results suggest that the extended Ly$\alpha$ emission is consistent with a cooling flow of accretion \citep{2011Stewart} -- the inflow deposits gas and angular momentum into the circumgalactic medium (CGM), transports cool gas to the galaxy, and maintains large star formation rates at high-redshift. However, observations of Ly$\alpha$ emission alone are unable to isolate whether the velocity shifts are due to radiative transfer effects or kinematics.  More observations using non-resonant lines (e.g. H$\alpha$, H$\beta$, etc.) alongside Ly$\alpha$ emissions suggest that there are variations of physical distance between emission regions and the associated quasar \citep{ouchi2020,arrigonietal19}.  These variations imply that the emission structure is oriented along our line of sight with its length extending out to Mpc scales.  The \citet{cantalupo19} Multi Unit Spectroscopic Explorer (MUSE) non-resonant line observations of the Slug Nebula are different from those of PCWI, since MUSE data shows abrupt velocity gradients in the nebula's neighboring regions.  This implies that there may be more complex kinematics than the rotational motion proposed by \citet{2015Martin}.  \citet{caietal17a} detected a giant LAB, MAMMOTH-1, extending to nearly $450$ kpc despite a significantly shallower observation than either the Jackpot or Slug nebulae. MAMMOTH-1, observed with narrow-band imaging, is $1.5\times$ the size of the Slug (z $\simeq$ 2.28) and Jackpot (z $\simeq$ 2.04) nebulae when measured from SB contours of $4.8\times10^{-18}$ ergs s$^{-1}$ cm$^{-2}$ arcsec$^{-2}$. MAMMOTH-1 resides in a unique field where there are several strong Ly$\alpha$ absorbers at $z=2.32\pm0.02$ which are associated with 5 background QSOs, projected within 20 $h^{-1}$ Mpc. The MAMMOTH survey has also detected a second LAB \citep{caietal17b}, indicating that these large LABs are well associated with clustered QSOs.

In addition to this evidence for a LAB connection with clustered QSOs, \citet{2016Borisova} have targeted ultra-luminous QSOs with $g<19$ at $z=3$ and find 100\% incidence of extended Ly$\alpha$ emission. 
\citet{caietal19} found extended Ly$\alpha$ emission associated with 14 out of their 16 QSO sample at $z=2.1-2.3$. Among those 14 detected Ly$\alpha$ nebulae, 4 had physical extents and luminosities consistent with ELANe. Results from the Fluorescent Lyman-Alpha Structures in High-z Environments (FLASHES) Survey \citep{osullivanetal20} indicates significant detections of Ly$\alpha$ emission around galaxies with 2.4 $<$ z $<$ 2.75, including several QSOs covered by the Keck Baryonic Structure Survey \citep{2012Rudie}.
With the Multi Unit Spectroscopic Explorer (MUSE), \citet{arrigonietal19} observed a total sample of 61 extended Ly$\alpha$ emission nebulae that have an average maximum projected distance of 80 kpc, each associated with quasars at $z=3.03-3.46$.  At higher redshifts, \citet{bielby2020} detected an extended Ly$\alpha$ nebula around a z$\sim$5.26 quasar. Additionally, in the Reionization Epoch QUasar InvEstigation with MUSE (REQUIEM) Survey, 12 of the 31 z$=5.7-6.6$ quasars were associated with Ly$\alpha$ nebulae \citep{farina19}. \citet{drake2019} confirmed the presence of extended Ly$\alpha$ halos from 4 out of 5 observations at z$\sim6$.  In light of these detections of extended gas around both clustered and single QSOs at a range of redshifts, we attempt to more systematically understand the extent and kinematics of gas around the most densely clustered QSOs at a slightly lower redshift, focusing on z $\simeq$ 2.4.  

Here, we describe our results from initial observations of our 4 survey targets using the Palomar Cosmic Web Imager (PCWI) at the 200 inch Hale Telescope. 

\section{Target Selection}\label{sec:Target}

We selected our targets from the SDSS-IV QSO database of 200,000 QSO spectra \citep{parisetal17}, providing a set of targets between $z=2.2-2.8$ that are accessible with PCWI. From this sample we select 12 QSO groups for a multi-year observing program. These fields were selected because they contain the strongest clustered QSO groups in 1' area (to roughly match the FoV of PCWI). The extreme over-dense nature of these fields is further suggested by strong IGM Ly$\alpha$ absorption in the spectrum of background QSOs on a larger scale of $\sim10\ h^{-1}$ Mpc \citep{caietal16a}. Of these 12 target QSO groups, only 4 were observed with reliable data to date (see Section \ref{sec:observations}). 

Our selection criteria involved finding QSO pairs with separations less than 1', redshifts within $\Delta$z=0.02 of each other, at least one QSO having g $<$ $18$ mag, and observable with CWI. In the first stage of our CWI program, we search for extended emission that bridge the QSO pairs and are in the periphery of the QSOs. Higher spectral resolution follow up observations with KCWI would provide higher signal to noise observations that illuminate details such as gas kinematics and metal emission lines. ELAN0101+0201 at z=2.4, the closest separation pair, is the first of our survey to be published by \citet{caietal18}. Details of the QSO pair sample are included in Table \ref{tab:qsos}.

\section{Observations}\label{sec:observations}

The Palomar Cosmic Web Imager is a slicer integral field unit (IFU), designed for observing low surface brightness, diffuse objects, in particular emission from the CGM \citep{2010Matuszweski}. CWI uses a $40\arcsec$ by $60\arcsec$ image slicer to divide the field into 24 slices with area $40\arcsec$ by $2\arcsec.5$, resulting in a sampling of $\sim0.55''$/pix. CWI uses a novel observing mode, nod-and-shuffle, which allows for near perfect sky subtraction.  
CWI is mounted on the Hale 5 m telescope Cassegrain focus, at Palomar Observatory. CWI, with a full suite of gratings, covers a bandpass from $380$ to $770$ nm, with the ability to observe a wavelength range of $45$ nm at a time. Our observations used the Richardson grating (R $=2000$) and blue filter. A more thorough description of the instrument, observing strategies, and data pipeline can be found in \citet{2014Martin}.

\begin{table*}[t]
\tiny
\centering 
\begin{tabular}{ c | c | c | c | c | c | c | c | c | c | c | c| c } 
\hline\hline 
Name & RA & DEC  & RA & DEC &  $z^a$ & $z^a$ & g-band Mag$^a$ & g-band Mag$^a$ & i-band Mag$^a$ & i-band Mag$^a$ & separ. & Central $\lambda$ \\ 
 & QSO$_1$ &QSO$_1$ &QSO$_2$ &QSO$_2$ &QSO$_1$ &QSO$_2$ & QSO$_1$ & QSO$_2$ &  QSO$_1$ & QSO$_2$ & arcsec & $\rm{\textrm{\AA}}$ \\ 

\hline
ELAN0101$^b$ & 01:01:16.5   &  +02:01:57.4   & 01:01:16.9   &  +02:01:49.8& 2.443 & 2.459 & 18.18$\pm$0.01 &21.68$\pm$0.06 & 18.19$\pm$0.01 & 21.62$\pm$0.11 & 9$\arcsec$ & 3920 \\
\hline
J1613 (T2) & 16:13:02.0 & +08:08:14.3 & 16:13:01.7 & +08:08:06.1 & 2.3864 & 2.3818 & 18.91$\pm$0.01 & 19.55$\pm$0.01 & 18.80$\pm$0.01 & 19.47$\pm$0.02 & 9.45$\arcsec$ & 4120 \\
\hline
J1120 (T3) & 11:20:53.2 & +46:33:35.4 & 11:20:54.2  &+46:33:27.2 & 2.5122 & 2.5108 & 18.62$\pm$0.01 & 21.87$\pm$0.06 & 18.60$\pm$0.01 & 21.67$\pm$0.12 & 13.17$\arcsec$ & 4270 \\
\hline
J1334 (T4) & 13:34:24.7 & +45:28:55.4 & 13:34:23.5 & +45:29:02.3 & 2.2620 & 2.2519 & 17.97$\pm$0.01 & 19.32$\pm$0.01 & 17.78$\pm$0.01 & 19.31$\pm$0.02 & 14.38$\arcsec$ & 3960 \\
\hline
J1342 (T5) & 13:42:10.8 & +60:35:22.3 & 13:42:06.1 & +60:15:06.1 & 2.3975 & 2.3850 & 17.93$\pm$0.01 & 21.33$\pm$0.04 & 21.97$\pm$0.20 & 21.45$\pm$0.09 & 47.9$\arcsec$ & 4120 \\
\hline

\end{tabular}
\caption{\label{tab:qsos} Table of QSO pair characteristics and observational parameters for our PCWI sample.  The seeing for observations varied between 1-2". \\
$^{a}$Values from SDSS DR17.\\
$^b$Results from \citet{caietal18}.
}
\end{table*}

\subsection{Data Reduction}\label{sec:datareduction}

We used the standard CWI pipeline to perform the data reduction \citep{Martin2014CWI_II}. The CWI data pipeline consists of a multi-stage process which yields a 3-D data cube from each image. We use the pipeline to build a master bias and flat field image (ideally from twilight flats) for each night of observations. Cosmic rays are then removed, and a geometric solution is derived to identify and extract slices, and to verify wavelength coverage across the field of view. For nod and shuffle observations, as in this case, the sky image is subtracted from the object image before further processing. A sky cube is also generated for verification and diagnostics. Slice to slice, or relative response, corrections are made based on the master flat and final standard star calibration is performed to convert the data cube pixel intensities to physical flux units. For our chosen observing mode, one data cube is generated, containing $\sim$20 minutes of sky-subtracted data on the target object. For multiple images of the same target, data cubes are registered by hand to ensure targets are spatially aligned, removing flexure or pointing errors as much as possible. 

Further data analysis outside of the CWI pipeline was done using \textit{CWITools}\footnote{\url{https://github.com/dbosul/cwitools}}, a software library developed by author Donal O'Sullivan. For each of our targets, we stacked all cubes, aligning along the bright central QSO. PSF subtraction is performed using the same methodology as \citet{osullivan2020} to subtract bright QSOs and isolate extended nebular emission.  First, a broadband image is created by summing over all the wavelength layers of the cropped cube.  DAOStarFinder from the package Photutils \citep{photutils2020} is used to identify QSOs in the white light image above a signal-to-noise ratio (SNR) threshold of $\sim5\sigma$ from the broadband image.   
For each QSO, a 3D PSF model is built and subtracted from the data cube.  In order to create a 3D PSF model, a 2D PSF model is made by integrating over  wavelength windows of 150$\textrm{\AA}$, centered on each wavelength channel.  Then, the 2D PSF model is scaled to match the PSF in each wavelength pixel  (0.55{\AA} wide).  The scaling factor is determined by minimizing the residual sum of squares within a radius of $\sim1".5$, and the scaled PSF is subtracted from each wavelength channel out to a radius of $\sim5"$, which is typically about twice the seeing.  We found a SNR threshold of $\sim5\sigma$ and a radius of $\sim5"$ sufficiently removed the bright region of PSFs.  Pixels used for scaling purposes are masked and excluded from future analysis.  Ly$\alpha$ wavelengths were not masked in building the empirical PSF model.  The analysis between masked and unmasked PSF models showed that the 5$"$ radius circular masks centered on PSF subtracted QSOs in the moment maps mask the entire area of the PSF subtraction and any differences that would arise between the two models.

Residual background is present due to small amounts of diffuse scattering from the camera, the grating, and light leakage under the nod-and-shuffle mask.  The background has no structure comparable to extended emission lines, but removal creates a cleaner smoothed image.  We therefore subtract the remaining background structure from these cubes using a first order polynomial fit.

The data cubes are then smoothed using an adaptive kernel at a SNR threshold of $\sim$3.5 to better outline faint diffuse emission.  They are smoothed using boxcar wavelength kernels and gaussian spatial kernels, first spectrally, then spatially.  First, data cubes are spectrally smoothed by 2$\textrm{\AA}$ kernels.  Then, spatial smoothing starts at a FWHM of 3 pixels followed by larger smoothing kernels in 2 pixel increments up to 20 pixels.  When a spaxel (spatial pixel that can move in a third dimension, the wavelength axis) exceeds the noise threshold, it is selected for the final cube as a detection.  Before the next spatial smoothing cycle, flux within the smoothed spaxel is subtracted from the unsmoothed cube.  Subsequently, this unsmoothed difference cube is spectrally smoothed by a 4$\textrm{\AA}$ kernel, afterwards, spatial smoothing is repeated.  This algorithm repeats a last time using an 8$\textrm{\AA}$ kernel for spectral smoothing.  This process also creates a SNR cube for each object that is used to define the 2$\sigma$ contours in Figures \ref{fig:T2}, \ref{fig:T3}, \ref{fig:T4}, and \ref{fig:T5} \citep{Martin2014CWI_I, Martin2014CWI_II, martin2015, Martin2016, Martin2019, osullivan2020_flashes}.

After smoothing, potential Ly$\alpha$ emission in the cube was found by segmenting objects in the data cube and creating an object cube.  During segmentation, voxels (0.594" x 0.594" x 0.55$\textrm{\AA}$) above a SNR of 4 were identified as detections, but only groups of 500 voxels or more were selected to create the object cube.  Finally, pseudo-narrow band (PNB) images are created by summing up all wavelength channels in the final unsmoothed data cube with the chosen objects from the object segmentation cube, and converting the image from flux (erg s$^{-1} $ \textrm{cm}$^{-2}$ \textrm{\AA}$^{-1}$) to surface brightness (erg s$^{-1} $ \textrm{cm}$^{-2}$ \textrm{arcsec}$^{-2}$) units.  First (velocity) and second (dispersion) moment maps are created from the PNB image.  One-dimensional spectra are made by summing along the two spatial axes for the specified object.  \citet{osullivan2020} details the process of creating these data products within CWITools.

\section{Results}

Extended Ly$\alpha$ emission was detected in all four targets, but only one target has characteristics consistent with the definition of an ELAN ($>$200kpc and $>$10$^{44}$erg s$^{-1}$).  The 1$\sigma_{\textrm{SB}}$ sensitivity of each target (Table \ref{tab:targets}) was calculated over a rest-frame velocity range within $\pm$500 km s$^{-1}$ of the Ly$\alpha$ line based on the mean redshift of each QSO pair.  An image was made for each target by integrating over wavelength channels of this velocity range.  At least fourteen 1arcsec$^{2}$ circular apertures were used on each $\pm$500 km s$^{-1}$ integrated image to calculate the 1$\sigma_{\textrm{SB}}$ sensitivity.  We characterize each nebula's physical extent by measuring the largest angular separation on the 2$\sigma$ central large surface brightness contour within the produced pseudo-narrowband Ly$\alpha$ images. Each target's total Ly$\alpha$ luminosity was measured by summing background-subtracted emission in the pseudo-narrowband Ly$\alpha$ images. Spectral resolution limits velocity dispersion measurements to a floor of $\sim$200 km/s.  Nebulae calculated star formation rates were derived from the H$\alpha$ star formation rate \citep{kennicutt98}.  In Table \ref{tab:targets}, it is assumed that star formation is due to case-B recombination, where the relation for H$\alpha$ and Ly$\alpha$ luminosity of the source is $L_{Ly_{\alpha}}$ = 8.7$L_{H_\alpha}$ \citep[e.g.,][ ]{dijkstra2010}.

\subsection{J1613 (T2)}
\begin{figure*}[t]
    \centering
    \includegraphics[width=\textwidth]{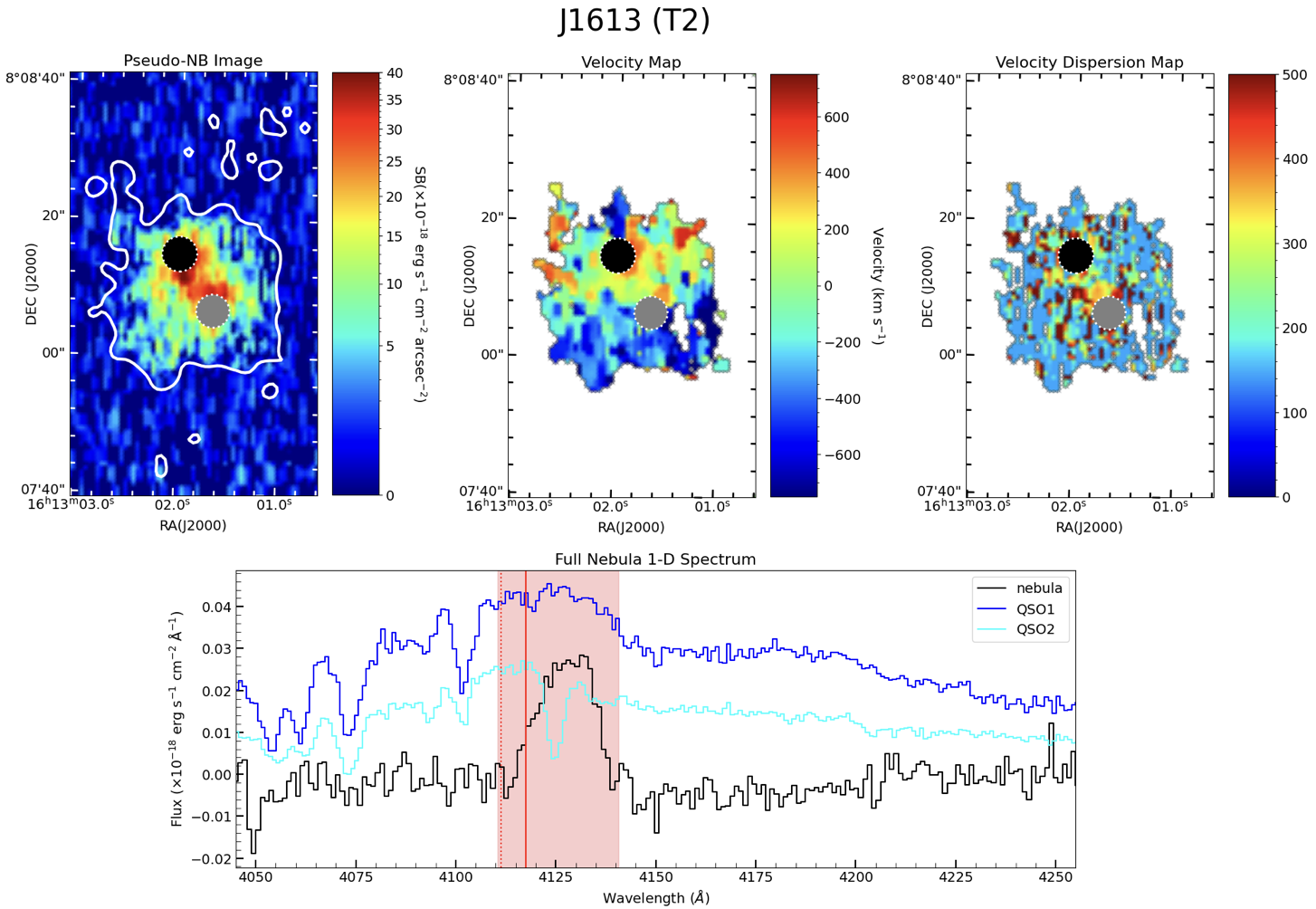}
    \caption{Target J1613 PCWI Ly$\alpha$ emission results. The upper left panel shows the pseudo-narrowband (PNB) Ly$\alpha$ image created from the wavelength range shown in the shaded red region in the bottom panel with the full nebula 1-D spectrum. Black (QSO 1) and gray (QSO 2) circles in the upper panels represent the positions and PSF sizes of the QSOs that were PSF subtracted out. White contours in the PNB image represent the calculated 2$\sigma$ Ly$\alpha$ emission surface brightness level.  The upper middle and upper right panels show the velocity map (moment 1), and velocity dispersion map (moment 2) respectively.  The moment maps use the average Ly$\alpha$ line emission of the QSO pair at z$\sim2.385$ as the rest-frame velocity of the system.  They are calculated based on data points within the central large 2$\sigma$ contour in the PNB image.  The smaller 2$\sigma$ contours are excluded since they appear to be regions of noise.  The PNB, velocity, and velocity dispersion maps are smoothed relative to adjacent pixels using gouraud shading.  The lower panel shows the full range of 1-D nebular spectrum plotted with the associated QSO pair. A red vertical solid (dotted) line is drawn in the 1-D spectrum at the wavelength of Ly$\alpha$ emission at the redshift of QSO 1 (QSO 2). The 1-D QSO 1 and QSO 2 spectra fluxes are scaled by $10^{-4}$ to fit on the same axis range as the nebular spectrum.  } 
    \label{fig:T2}
\end{figure*}

We present the PCWI PSF-subtracted Ly$\alpha$ cube products for target J1613 in Figure \ref{fig:T2}. The 6600s total exposure time on target J1613 yields an empirical 1$\sigma_{SB}$ of 8.69$\times$10$^{-18}$ erg s$^{-1}$ cm$^{-2}$ arcsec$^{-2}$ within a wavelength range of 4107.72 to 4122.57$\textrm{\AA}$.  The PNB image contains wavelengths 4110.47 to 4140.72$\textrm{\AA}$, and the 2$\sigma$ contour encompasses an area of 576 arcsec$^{2}$.  Ly$\alpha$ emission in the PNB has an angular extent of $37".4$, corresponding to a projected physical size of 305 kpc at the mean redshifts of QSO 1 and 2. The total Ly$\alpha$ luminosity associated with target J1613 is 1.80$\times$10$^{44}$ erg s$^{-1}$. The extended size and Ly$\alpha$ luminosity both sufficiently exceed the criteria for ELAN classification. The data depth, extent, and Ly$\alpha$ luminosity for all sample targets are displayed in Table \ref{tab:targets}.

Like the quasar pair seen in \citet{battaia2019b} where there is an intergalactic bridge connecting two QSOs, the Ly$\alpha$ emission morphology resembles structure extending between the two QSOs.  The velocity gradient across where the J1613 QSO pair resides may indicate interaction.  The velocity dispersion is limited by the spectral resolution of the instrument $\lesssim$ 200 km s$^{-1}$, however dispersions greater than that appear between the QSOs and around both QSOs.  Shock heating and photoionization are potential powering mechanisms for the ELANe in the CGM of this system (see Section \ref{poweringmechanisms}).

\subsection{J1120 (T3)}
\begin{figure*}[t]
    \begin{center}
    \includegraphics[width=\textwidth]{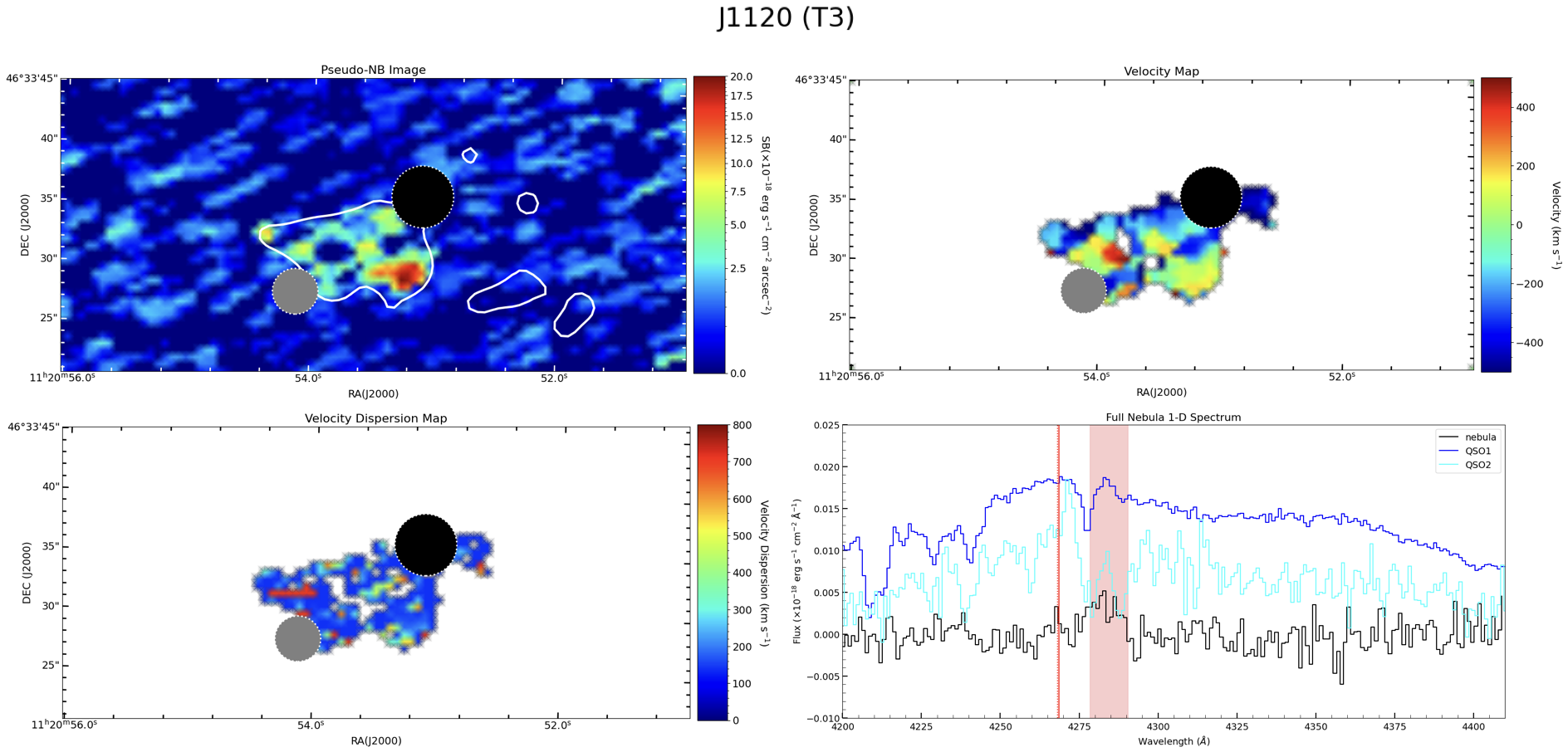}
    \caption{Target J1120 PCWI Ly$\alpha$ emission results. The upper left panel shows the PNB Ly$\alpha$ image created the wavelength range shown in the shaded red region in the bottom right panel of the full nebular 1-D spectrum. The upper right and lower left panels show the velocity and velocity dispersion maps respectively.  These moment maps use the average redshifted Ly$\alpha$ emission line of the QSO pair at z$\sim2.51$ as the rest-frame velocity of the system. The 1-D QSO1 and QSO2 spectra fluxes are scaled by $10^{-2.5}$ and $10^{-1.5}$ respectively to fit on the same axis range as the nebular spectrum. The panel display characteristics are mirrored from Figure \ref{fig:T2}.}  
    \label{fig:T3}
    \end{center}
\end{figure*}

The data products for target J1120 are shown in Figure \ref{fig:T3}. The 3600s total exposure time yields a 1$\sigma_{SB}$ of 8.92$\times$10$^{-18}$ erg s$^{-1}$ cm$^{-2}$ arcsec$^{-2}$ within a wavelength range of 4261.33 to 4275.63$\textrm{\AA}$.   The PNB image contains wavelengths 4278.38 to 4290.48 $\textrm{\AA}$, and the 2$\sigma$ contour encompasses an area of 62 arcsec$^{2}$. The Ly$\alpha$ emission has an angular extent of $20".2$, corresponding to a projected physical size of 163 kpc at the average redshifts of QSO 1 and 2. The total Ly$\alpha$ luminosity associated with target J1120 is 0.11$\times$10$^{44}$ erg s$^{-1}$, which does not meet the criteria for ELAN classification.  Future Lya observations of J1120 that probe to deeper surface brightness levels may uncover additional Lya emission.
The kinematic distribution shows a bulk of the gas flows near the systemic velocity. 

There is a bulk of blueshifted features (negative velocity in the velocity map) on the northernmost edge of the Ly$\alpha$ emission region. It is possible that this feature is a shock front propagating outward from a cluster galaxy within the nebula. The low average velocity dispersion is potentially caused by the relative lack of mechanical energy input from feedback. 

\subsection{J1334 (T4)}

\begin{figure*}[b]
    \begin{center}
    \includegraphics[width=\textwidth]{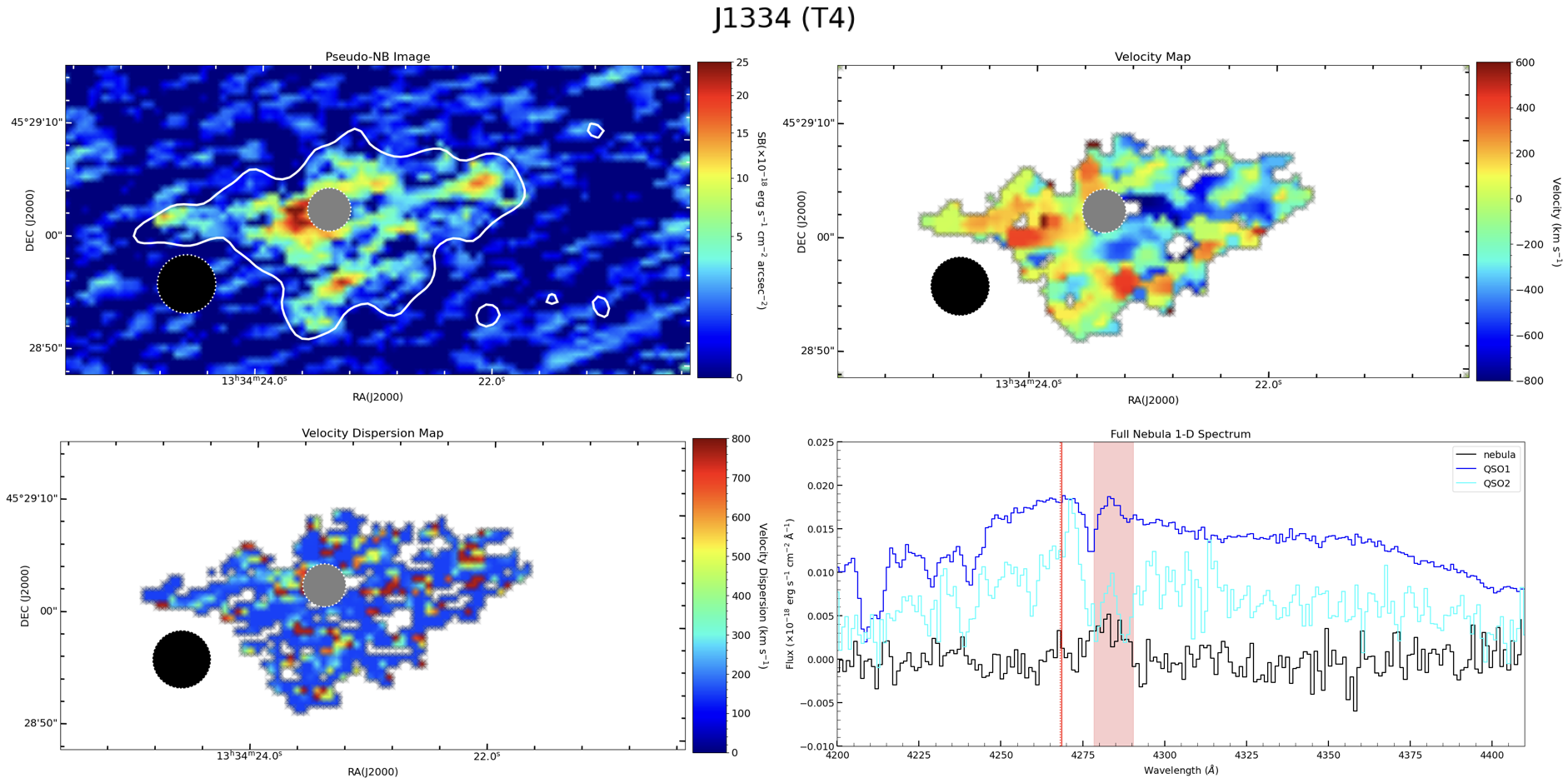}
    \caption{Target J1334 PCWI Ly$\alpha$ emission results. Velocity and velocity dispersion maps use the average Ly$\alpha$ emission line of the QSO pair at z$\sim2.257$ as the rest-frame velocity of the system. The 1-D QSO 1 and QSO 2 spectra fluxes are scaled by $10^{-2.5}$ and $10^{-2}$ respectively to fit on the same axis range as the nebular spectrum. The panel layout and display characteristics are mirrored from Figure \ref{fig:T3}.}
    \label{fig:T4}
    \end{center}
\end{figure*}

The data products for target J1334 are shown in Figure \ref{fig:T4}. The 6000s total exposure time yields a 1$\sigma_{SB}$ of 8.58$\times$10$^{-18}$ erg s$^{-1}$ cm$^{-2}$ arcsec$^{-2}$ within a wavelength range of 3952.31 to 3965.5 $\textrm{\AA}$.  The PNB image contains wavelengths 3967.71 to 3988.06 $\textrm{\AA}$, and the 2$\sigma$ contour encompasses an area of 311 arcsec$^{2}$. The Ly$\alpha$ emission has an angular extent of $34".5$ corresponding to a projected physical size of 284 kpc at the average redshifts of QSO 1 and 2. The total Ly$\alpha$ luminosity associated with target J1334 is  $0.40 \times10^{44}$ erg s$^{-1}$, which does not meet the criteria for ELANe classification.  However, the structure of J1334 is remarkably clumpy, so the source may host numerous Ly$\alpha$ emitting galaxies (LAEs). Future KCWI observations with higher spatial and spectral resolution can make identifying individual sources possible. 

\subsection{J1342 (T5)}

\begin{figure*}
    \begin{center}
    \includegraphics[width=\textwidth]{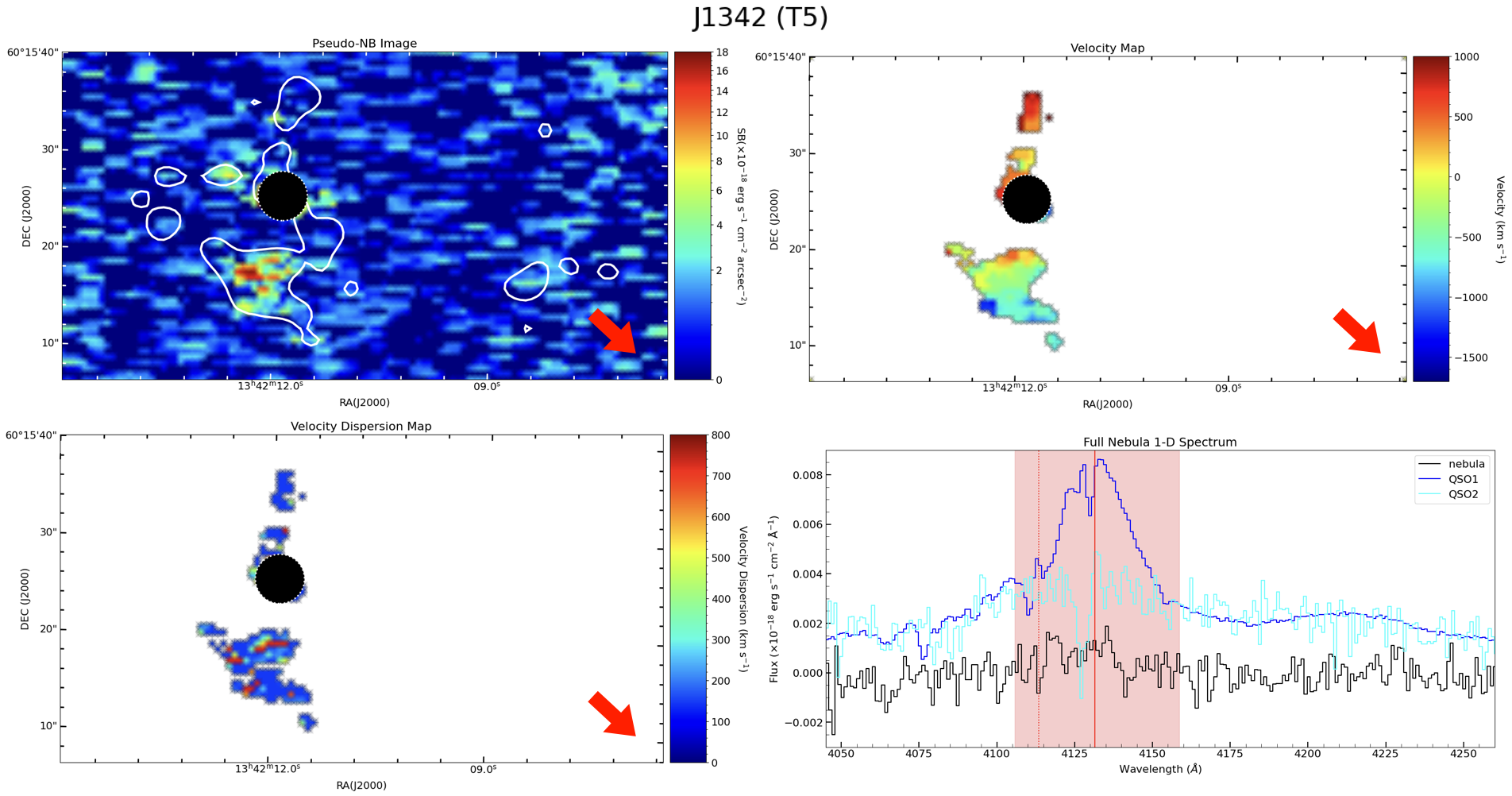}
    \caption{Target J1342 PCWI Ly$\alpha$ emission results. The moment maps use the average Ly$\alpha$ emission line of the QSO pair at z$\sim2.39$ as the rest-frame velocity. The 1-D QSO 1 and QSO 2 spectra fluxes are scaled by $10^{-3.5}$ and $10^{-2}$ respectively to fit on the same axis range as the nebular spectrum. The red arrows are in the direction of the location of QSO 2 outside of the field of view. The panel layout and display characteristics are mirrored from Figure \ref{fig:T3}.}
    \label{fig:T5}
    \end{center}
\end{figure*}

The data products for target J1342 are shown in Figure \ref{fig:T5}. The 7200s total exposure time yields a 1$\sigma_{SB}$ of  8.39$\times$10$^{-18}$ erg s$^{-1}$ cm$^{-2}$ arcsec$^{-2}$ within a wavelength range of 4115.74 to 4129.49  $\textrm{\AA}$.  The PNB image contains wavelengths 4105.84 to 4158.64 $\textrm{\AA}$, and the 2$\sigma$ contour encompasses an area of 121 arcsec$^{2}$. The Ly$\alpha$ emission has an angular extent of $27".5$ at the average redshifts of QSO 1 and 2. The total Ly$\alpha$ luminosity associated with target J1342 is $0.37\times10^{44}$ erg s$^{-1}$. The Ly$\alpha$ luminosity does not meet the criteria for ELANe classification.
Features surrounding the main central 2$\sigma$ contour were masked in the calculation of velocity and velocity dispersion maps to exclude artifacts that resemble noise.  The north to south velocity gradient suggests the presence of a bipolar flow from QSO 1.

\begin{figure*}
    \begin{center}
    \includegraphics[scale=0.5]{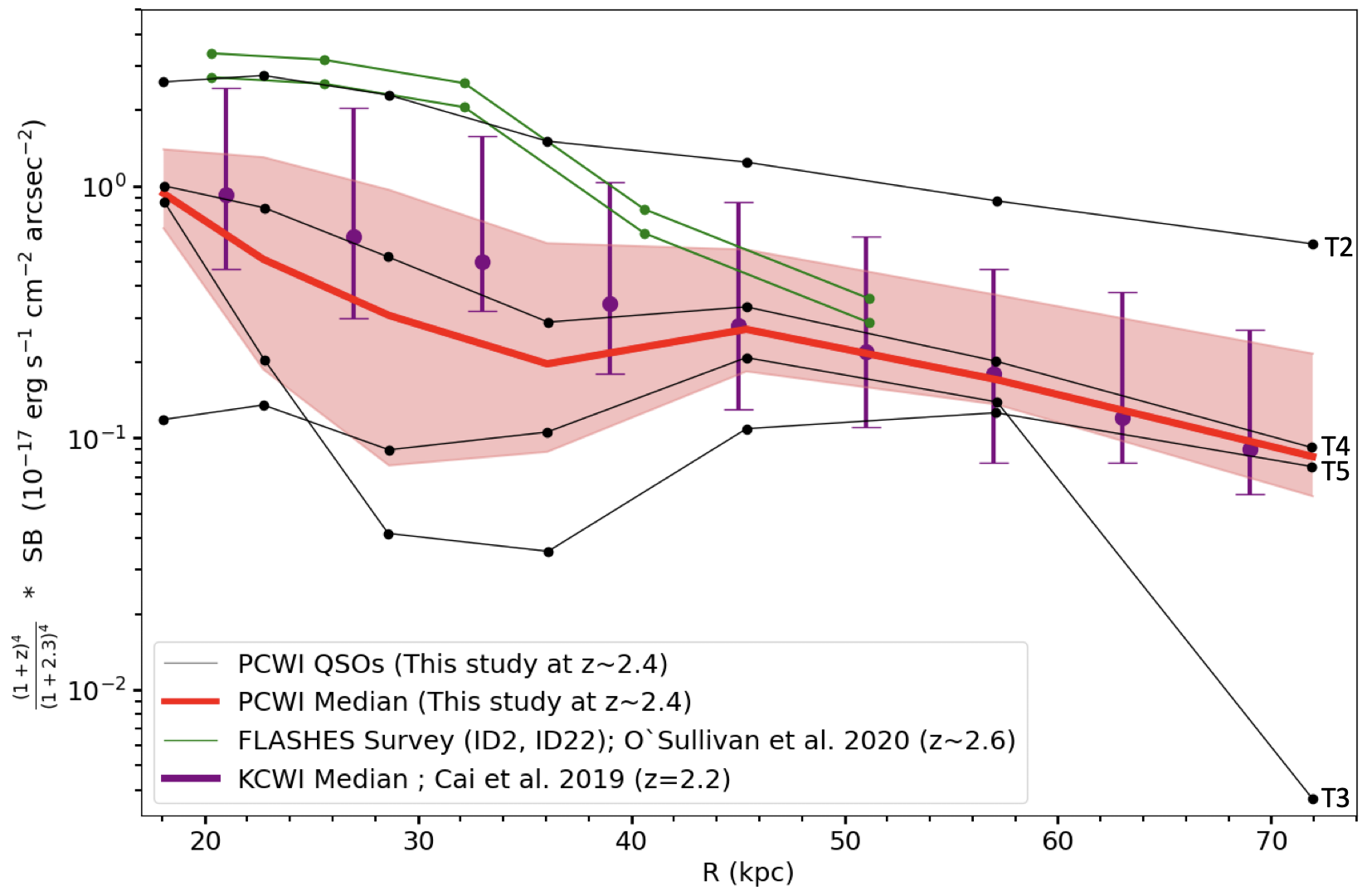}
    \caption{Ly$\alpha$ surface brightness (SB) profiles as a function of radius from the brighter QSO of the QSO pair. These SB profiles were corrected for cosmological dimming.  Profiles were created from averaging the SB within annuli defined at increasing distances from the brighter QSO.  The black lines are the SB profiles from the current PCWI study, each target is associated with a QSO pair at z$\sim$2.2-2.5 (Table \ref{tab:targets}).  The red line is the median, and the red shaded region represents the SB within 25 to 75 percentiles of this study.  Purple data points represent the median, and their error bars show the 25 to 75 percentile of the SB profile from the \citet{caietal19} KCWI sample of 16 QSOs at z=2.1-2.3.  Green data points represent radial profiles around two individual QSOs, a bright and an intermediate Ly$\alpha$ emission detection from the FLASHES survey \citep{osullivan2020_flashes}.  Detection drops to around zero for this study at $\gtrsim$ 50 kpc.}
    \label{SB_plot}
    \end{center}
\end{figure*}

\begin{table*}[!hb]
\centering 
\begin{tabular}{ c | c | c || c | c | c | c} 
\hline\hline 
Target & t exp & 1$\sigma_{SB}$ ($\times10^{-18}$ & Angular Extent & L$_{\rm{Ly}\alpha}$ & $^a$SFR$_{\rm{Ly}\alpha}$ & ELANe?\\ 
Name & (sec) & erg s$^{-1}$ cm$^{-2}$ arcsec$^{-2}$) & (kpc)& ($\times10^{44}$ erg s$^{-1}$) & (M$_{\odot}$ yr$^{-1}$) &  \\
\hline 
ELAN0101$^b$ & 14400 & 4.50 & $232$ & 4.5 & 409 & Yes\\
J1613 (T2) & 6600 & 8.69 & $305$ & 1.80 & 163 & Yes\\
J1120 (T3) & 3600 & 8.92 & $163$ & 0.11 & 10  & No\\
J1334 (T4) & 6000 & 8.58 & $284$ & 0.40 & 36 & No \\
J1342 (T5) & 7200 & 8.39 & $224$ & 0.19 & 17 &  No\\
\hline
\end{tabular}
\caption{\label{tab:targets} Data depth and Ly$\alpha$ nebulae characteristics. Calculations for 1$\sigma_{SB}$ are based off of 5$\sigma$ contour measurements.  Angular extents are maximum extents} on the largest central 2$\sigma$ surface brightness contour of each target's FoV.  Distance calculations are based on a $\Lambda$CDM cosmology with $\Omega_m = 0.3$, $\Omega_\Lambda = 0.7$, and $h = 0.7$ \citep{Wright2006}. Ly$\alpha$ luminosity and SFR are calculated for emission regions within 2$\sigma$ detection.\\
$^a$Calibration from \citet{kennicutt98} and \cite{dijkstra2010}.
\end{table*}

\section{Discussion}
Ly$\alpha$ emission was found around all QSO pair targets, but only one reaches the criteria of an ELANe, this target (J1613) has a Ly$\alpha$ luminosity $>$ 10$^{44}$ erg s$^{-1}$ and a projected physical size $>300$ kpc. These enormous and extremely bright nebulae are exceedingly rare near $z\sim2$. Below, we explore relationships between these nebulae and their host QSOs, as well as their powering mechanisms. 

\subsection{Integrated Sample Properties with QSO Characteristics}

The Ly$\alpha$ nebulae sizes and integrated Ly$\alpha$ luminosities are shown in Table \ref{tab:targets}. We explore correlations of these results with properties of both QSO 1 and QSO 2 for each target. In our data, there is no observed correlation between L$_{\rm{Ly}\alpha}$ and QSO magnitude in any of the SDSS bands, or between Ly$\alpha$ nebula size and QSO magnitude. 
Lastly, there is no redshift evolution over the range of system redshifts in our sample with neither L$_{\rm{Ly}\alpha}$ nor Ly$\alpha$ nebula size. 

\subsection{ELANe Powering Mechanisms}\label{poweringmechanisms}

We follow the approach of \citet{caietal18} to explore the powering mechanisms of the nebulae in our new sample. Extended Ly$\alpha$ emission can be powered by shock heating from AGN outflows, photoionization, gravitational cooling, and resonant scattering from the QSO broadline region \citep{cantalupo17}.  First, AGN outflows are expected to behave like fast shocks which ionize the material around quasars, resulting in Ly$\alpha$ emission where $F_{Ly\alpha}$ $\propto$ $n_H$$v_{\rm{shock}}^3$ (e.g., \citet{allen2008}). Second, Ly$\alpha$ emission could be produced by recombinations following photoionization of hydrogen by the hard spectrum of quasars.  In this scenario, as we discuss below, there are two limiting cases for optically thin and optically thick gas due to ionizing radiation (e.g., \citet{hennawietal13}).  Ly$\alpha$ fluorescence on the edges of optically thick clouds arises where hydrogen is photoionized by the UV radiation of quasars.  Within these clouds, resonant scattering of Ly$\alpha$ photons experience absorption and re-emission by ground state hydrogen atoms until they finally escape the cloud \citep{Cantalupo2005, 2013prochaska}.  Lastly, Ly$\alpha$ emission could be due to gravitational cooling as  collisionally excited neutral hydrogen at $\gtrsim$ 10$^{4-5}$ K flows into galaxies \citep{geach2016,steideletal11,dijkstra2009}.   

Since kinematics show evidence of potential AGN outflows and that nebulae may be photoionized by one or both QSOs in each QSO pair, the subsequent analysis will only focus on the former two scenarios.  AGN outflows eject gas symmetrically outward, shock heating surrounding gas. These motions could manifest as a bipolar kinematic pattern the nebular emission velocity maps, depending upon the orientation of the outflow feature. Two of our sample targets, J1613 (T2) and J1342 (T5), have kinematic patterns that may indicate the presence the bipolar flows. \citet{alexanderetal10} and \citet{harrisonetal14} suggest that AGN outflows can eject matter at velocities upward of $v_{\rm{max}}\gtrsim1500$ km s$^{-1}$. Though the velocity maps of ELANe J1613 (T2) and J1342 (T5) do not show velocities greater than $1500$ km s$^{-1}$, the coherent velocity structures reach larger extrema than our two other sample targets. Furthermore, an outflow oriented slightly away from our line of sight may lead to an underestimate of the absolute velocity of the gas due to projection.  However, improved higher resolution data is needed to definitively determine whether or not these targets have bipolar flows.

To explore the photoionization scenario, we assume that a large majority of ionizing photons come from QSO 1, and hence the nebulae are photoionized by QSO 1, except for J1613 where the QSOs are similar in brightness. \citet{hennawietal13} find the surface brightness of an optically thin cloud due to recombination follows:

\begin{align*}
    {\rm{SB}^{\rm{thin}}_{\rm{Ly}\alpha} = 8.8\times10^{-20} \left(\frac{1+z}{3.253} \right)^{-4}\times \left(\frac{f^{\rm{thin}}_{\rm{C}}}{0.5} \right)} \times \\ \left(\frac{N_{\rm{H}}\times{n_{\rm{H}}}}{10^{20.5}\rm{cm}^{-5}}\right)\rm{erg}\rm{s}^{-1} \rm{cm}^{-2}\rm{arcsec}^{-2}
\end{align*} 
	\label{eq:1}

For this calculation, we assume values of covering factor, hydrogen column density, and hydrogen volume density.  Covering factor, is $f_{\rm{C}}^{\rm{thin}}=0.5$, as suggested by \citet{arrigonietal15a, arrigonietal15b, 2015Hennawi} for a relatively smooth Ly$\alpha$ morphology.  Hydrogen column density is approximately $N_{\rm{H}}\approx 1\times10^{20.5\pm1.0}$ cm$^{-2}$ based on photoionization models by \citet{lauetal16} used to estimate hydrogen ionization fraction, where they showed that hydrogen column density did not vary significantly up to a distance of 200 kpc from the QSO.  Hydrogen number density is taken to be n$_{\textrm{H}}$ $\approx 1$ cm$^{-3}$ for cold T$\sim$10$^{4}$ K gas in the CGM \citep{2013Tumlinson}, where the SB$_{\textrm{Ly}\alpha}\sim$ 10$^{-19}$ to 10$^{-17}$ erg s$^{-1}$cm$^{-2}$arcsec$^{-2}$ at $\sim$ 50 kpc of QSO 1 for all targets. The expected surface brightness of an optically thin cloud due to recombination is then $\sim$8$\times$10$^{-20}$ erg s$^{-1}$ cm$^{-2}$ arcsec$^{-2}$. This value is below the observed surface brightness of our sample within 200 kpc of QSO 1, which suggests that the CGM density of our extended Ly$\alpha$ nebulae sample is $\sim$2-3 orders of magnitude higher than nebulae found in more isolated QSOs at $z\sim2$ \citep{arrigonietal16}.  The same is also found with nebulae from the \cite{caietal19} KCWI study and FLASHES Survey which have SB$\sim10^{-17}$ erg s$^{-1}$ cm$^{-2}$ arcsec$^{-2}$. For the optically thick case \citep{hennawietal13, caietal18}:

\begin{align*}\label{eq:2}
\rm{SB}^{\rm{thick}}_{\rm{Ly}\alpha}=5.3\times10^{-17}\left( \frac{1+z}{3.45} \right) ^{-4}\times \left(\frac{f^{\rm{thick}}_{\rm{C}}}{0.5} \right)\times\\ \left(\frac{\rm{R}}{160 \rm{kpc}}\right)^{-2} \times \left(\frac{L_{\nu_{\rm{LL}}}}{10^{30.9}\rm{erg}\rm{s}^{-1}\rm{Hz}^{-1}}\right)\rm{erg} \rm{s}^{-1} \rm{cm}^{-2} \rm{arcsec}^{-2}
\end{align*}
	\label{eq:2}
 
For this estimation, we assume $f^{\rm{thick}}_{\rm{C}}=0.5$. \citet{caietal18} estimates Log $\frac{\rm{L}_{\nu_{\rm{LL}}}}{\rm{erg}\rm{s}^{-1}}=31.1$, by scaling the composite spectrum of QSO 1 to match the i-band magnitude of QSO 1 in ELAN0101, where $\rm{L}_{\nu_{\rm{LL}}}$ is the specific luminosity at the Lyman edge \citep{arrigonietal15b,lussoetal15}. The predicted surface brightness from the optically thick case at $\sim50$ kpc from QSO 1 is SB$^{\rm{thick}}_{\rm{Ly}\alpha}$ $\sim$ 10$^{-15}$ erg s$^{-1}$cm$^{-2}$arcsec$^{-2}$ which is $\sim$ 2 orders of magnitude higher than our SB$_{\textrm{Ly}\alpha}$.  This suggests that the covering factor of the optically thick gas may be lower than what is assumed for this estimation, or photoionization of optically thick gas is not the mechanism that significantly contributes to the observed Ly$\alpha$ emission.

\subsection{Surface Brightness Profiles of Ly$\alpha$ Emission}

Redshift-corrected circularly-averaged surface brightness profiles, are shown in Figure \ref{SB_plot}. A characteristic surface brightness of $\sim10^{-17}$ erg s$^{-1}$ cm$^{-2}$ arcsec$^{-2}$ is detected in the QSO pair samples at z $\approx$ 2.3. The chosen lower radial limit for all targets is 18.1 kpc from the brighter QSO of each QSO pair.
Surface brightness is calculated in increments of 10$^{log(r)+0.1}$ kpc, where $r$ is radial distance from the brighter QSO in kpc. These radial increments are the midlines of annuli with $\pm$10$^{log(r)+0.05}$ kpc thickness.  Each data point is the average surface brightness within a spatial annulus calculated from the targets' pseudo-narrowband images. The red line shows the median of our PCWI sample and the red region spans between the 25th to 75th SB percentile of the sample at incremental distances from QSO1. Radial profiles are plotted from $\sim$15 to 75 kpc, following the \citet{caietal19} study that a power law profile of Ly$\alpha$ SB centered at QSO 1 is only valid at this range for QSOs at z$\approx$2.3.  QSO 2 is beyond 75 kpc from QSO 1 for all targets.
These nebulae are likely in fairly disturbed environments and may be drawing their energy from multiple external sources. Hence, their surface brightness profiles may be more affected by their environments than Ly$\alpha$ nebulae identified through other means. Despite this, these profiles provide a valid comparison to other samples of LABs and ELANe.  We compare the results of Figure \ref{SB_plot} to surface brightness profiles of other Ly$\alpha$ nebulae at various redshifts and environments.  \citet{arrigonietal16} found 7 out of a sample of 15 QSOs at z$\sim$2 have Ly$\alpha$ emission within $\sim$50 kpc of the QSO.  At z$\sim$2.65, an analysis of 92 UV-continuum selected galaxies showed a presence of Ly$\alpha$ emission in 55$\%$ of the sample, where the Ly$\alpha$ nebulae had an extent of at least $\sim$80 kpc \citep{steideletal11}.  Ly$\alpha$ nebulae in both studies had an average SB$_{\textrm{Ly}\alpha}\sim$10$^{-17}$erg s$^{-1}$ cm$^{-2}$ arcsec$^{-2}$.  Higher redshift studies at z$\sim3-6$ by \citep{wisotzskietal16} detected Ly$\alpha$ emission in 21 of 26 galaxies, and compared with Ly$\alpha$ halos at z$\sim$0 concluded that higher redshift galaxies have Ly$\alpha$ regions that are 5$\times$ more extended with SBs of 10$^{-18}$ to 10$^{-17}$ erg s$^{-1}$ cm$^{-2}$ arcsec$^{-2}$.  At z$\sim3.17$, a MUSE sample of 61 extended Ly$\alpha$ nebulae around QSOs showed that most of the Ly$\alpha$ nebulosities' SB ($\gtrsim$10$^{-18}$erg s$^{-1}$ cm$^{-2}$ arcsec$^{-2}$) reside within 50kpc of the host QSO, regardless of environmental variants such as radio-loudness and the presence of active QSO companions \citep{arrigonietal19}.  A smaller MUSE study of 17 QSOs at z$\sim3.5$ found that every QSO is surrounded by giant Ly$\alpha$ nebulae of projected sizes ranging from $\gtrsim$100 to 320 kpc. Therefore, \citet{borisovaetal16} suggests that Ly$\alpha$ emission regions are more readily found around bright QSOs at z$\sim$3-4, in contrast to the low detection rate of Ly$\alpha$ nebulae around QSOs at z$\sim$2.  The bulk SB of Ly$\alpha$ nebulosities around QSOs seem to mostly reside within $<$50 kpc of its host QSO at all redshifts for all these studies.

We include the median of the sample from \citet{caietal19} as purple data points with error bars spanning between 25th to 75th percentile of the sample in Figure \ref{SB_plot}. Figure 4 of \citet{caietal19} shows surface brightness profiles of the composite of these studies, and we include a redshift-corrected y-axis in Figure \ref{SB_plot} for direct comparison to that study.  Shown in green are examples of a bright (z=2.49) and an intermediate (z=2.79) detection of quasar fields from the FLASHES (Fluorescent Lyman-Alpha Structures in High-z Environments) Survey observed with PCWI.  The FLASHES Survey consists of 48 quasar fields at an average z$\sim$2.6 and the redshift dimming corrected characteristic SB for the sample is $\sim$6$\times$ 10$^{-18}$ erg s$^{-1}$ cm$^{-2}$ arcsec$^{-2}$ \citep{osullivan2020_flashes}. These comparisons show relative agreement with the $z\sim2$ results of \citet{caietal19}, since the surface brightness values of our sample are generally slightly lower overall than those at $z\sim3$. As discussed in that study, this redshift evolution can be explained through two possibilities: either nebular emission at $z\sim2$ has less circular morphology or covering fraction than that at $z\sim3$, or the nebular surface brightness at $z\sim2$ is intrinsically fainter. This would imply a lower local density or gas mass at $z\sim2$ than at $z\sim3$.  However, a larger sample size of Ly$\alpha$ nebulae at z $\sim$ 2 would be needed to confirm whether or not there is strong evidence of redshift evolution between z $\sim$ 2-3.












\subsection{Detection Frequency in QSO Pair Systems and Future Observations}

The discovery of ELANe around QSO pair systems represents an opportunity to efficiently study the extreme CGM.  To date, nearly all QSO pair systems observed with PCWI/KCWI show evidence of extended Ly$\alpha$ emission (\citealt{caietal18} and this study). Future surveys of the CGM could use this success as a marker for target selection that would lead to greater detection confidence. QSO pair systems trace overdense regions that are likely progenitors to high mass clusters at $z=0$ \citep{onoueetal18}.  
Only one ELANe has been discovered in a sample of 61 QSOs at $z\sim3$ \citep{arrigonietal19}, representing a $\sim1\%$ probability of detection in a given QSO system. More ELANe can be discovered by specifically targeting multiple QSO systems at $z\sim3$ and beyond, like the system found by \citep{arrigonietal18}, where Ly$\alpha$ emission was associated with 2 Ly$\alpha$ emitters and 2 QSOs, and another system where Ly$\alpha$ emitting intergalactic structures form bridges between a QSO pair  \citep{battaia2019b}.
The need for higher resolution follow-up observations of these targets is great. The higher spatial resolution of KCWI would grant the ability to study individual substructures in the ELANe, such as LAEs and bright nodes in detail. The larger spectral resolution would allow for robust Ly$\alpha$ line profile fitting within individual substructures. Such an analysis could reveal asymmetries and broad/narrow line components. Line asymmetries are more accurate indicators of ISM outflows or line of sight absorption characteristics. The characteristics of broad/narrow line components could also reveal the existence of AGN in nebular substructures \citep{trainoretal16}. 

\section{Conclusion}

The discovery of ELANe around QSO pair systems at cosmic noon represents a unique opportunity to study the CGM without the need for entirely blind surveys. We detect extended Ly$\alpha$ emission in the PCWI data from all four sample targets between $2.26<z<2.51$. One of the four targets had intense Ly$\alpha$ emission consistent with the definition of an ELAN. We classify this emission region as ELAN, based on the threshold definition in Ly$\alpha$ extent and luminosity. To date, many QSO pair systems observed with PCWI/KCWI have shown extended and intense Ly$\alpha$ emission. We find that ELANe J1613 is possibly powered by AGN outflows, as inferred by the kinematic distribution of the nebular emission. Also, the circularly-averaged surface brightness of our targets are slightly lower than that from other studies of nebular emission at $z\sim3$. This may imply that there is a lower local density or gas mass at $z\sim2$ than at $z\sim3$. KCWI would grant the ability to study individual substructures in the ELANe, such as LAEs and bright nodes in detail. 

\section{Acknowledgements}
Resources supporting this work were provided by NSF AAG Grant 1716907 and the California Institute of Technology.  ZC acknowledges National Key R\&D Program of China (grant no.\ 2018YFA0404503), the National Science Foundation of China (grant no.\ 12073014). The science research grants from the China Manned Space Project with No. CMS-CSST2021-A05. 

\facility{Palomar}

\bibliography{QSOpairs.bib}
\bibliographystyle{apj}

\end{document}